\documentclass{PoS}

\usepackage{amsmath,amsthm,amssymb}
\usepackage{braket}
\usepackage{dsfont}
\usepackage{feynmp}
\usepackage{graphicx}
\usepackage[labelsep=quad,indention=10pt]{subfig} 
\usepackage{tikz}

\definecolor{sky}{rgb}{0.4,0.6,1.}
\definecolor{darkgreen1}{rgb}{0, .35, 0}

\title{Exclusive Channel Study of the Muon HVP}

\ShortTitle{Exclusive Channel Study of the Muon HVP}

\author{Mattia Bruno\\
        Theoretical Physics Department,\\
        CERN, 1211 Geneve 23, Switzerland\\
        E-mail: \email{mattia.bruno@cern.ch}}

\author{Taku Izubuchi\\
        Department of Physics and RIKEN-BNL Research Center,\\
          Brookhaven National Laboratory, Upton, NY, 11973, USA\\
        E-mail: \email{izubuchi@bnl.gov}}

\author{Christoph Lehner\\
        University of Regensburg, 93040 Regensburg, Germany, and\\
        Department of Physics,\\
          Brookhaven National Laboratory, Upton, NY, 11973, USA\\
        E-mail: \email{clehner@quark.phy.bnl.gov}}

\author{\speaker{Aaron S. Meyer}\\
        Department of Physics,\\
          Brookhaven National Laboratory, Upton, NY, 11973, USA\\
        E-mail: \email{ameyer@quark.phy.bnl.gov}}

\abstract{ The Hadronic Vacuum Polarization (HVP) is a dominant
  contribution to the theoretical uncertainty of the muon anomalous
  magnetic moment.  The uncertainty in a lattice QCD calculation of
  the connected light-quark contribution to the HVP is dominated by
  the long-distance region of the vector correlation function.
  Explicit studies of the exclusive channels of the HVP diagram make
  it possible to reconstruct the long-distance behavior of the
  correlation function.  This removes most of the statistical
  uncertainty of the correlation function.  In these proceedings,
  preliminary results of an exclusive study of the isospin symmetric
  connected-only vector-vector correlation function using a hybrid of
  distillation and A2A techniques are presented.  The computation is performed
  on 2+1 flavor M\"obius Domain Wall Fermion ensembles with physical
  pion mass.  Reconstruction of the long-distance correlation function
  will enable lattice-only calculations of the HVP to achieve
  precision similar to estimates of the HVP from the R-ratio method on
  the timescale of the new experimental measurements of the muon
  anomalous magnetic moment.  }

\FullConference{37th International Symposium on Lattice Field Theory - Lattice2019\\
                16-22 June 2019\\
                Wuhan, China}

\begin{document}
\DeclareGraphicsRule{*}{mps}{*}{}

\section{Introduction}
\label{sec:intro}

The Hadronic Vacuum Polarization (HVP) contribution to the muon anomalous magnetic moment
 is currently one of the leading contributions to the uncertainty in the theory error budget
 and is also one of the most difficult to estimate.
The muon $g-2$ is a precision experiment and is sensitive to physics beyond the Standard Model.
Currently, the experimental results from the BNL $g-2$~\cite{Bennett:2006fi}
 experiment exhibit a $3.7\sigma$ tension with theory predictions,
 which hints at possible new physics contributions.
With the upcoming release of first results from the Fermilab $g-2$ experiment,
 it is important to have a robust understanding of the current status of the theory
 prediction for the HVP.

The most precise determinations of the HVP come from studies using electron scattering data
 and the optical theorem to compute the spectral density, known as the
 R-ratio, see for example Ref.~\cite{Keshavarzi:2018mgv}.
However, this technique relies on experimental input for the spectral density for which tensions exist around the
 rho resonance peak, see, e.g., ~\cite{Davier:2019can}.  This poses a challenge to properly
quantify the uncertainty for the HVP contribution from the R-ratio method.



Lattice QCD (LQCD) is the only first principles method to access $a_\mu^{HVP}$
 and provides way to avoid or reduce the use of the R-ratio data.
Though LQCD calculations are not as precise as the R-ratio at this time,
 the LQCD determinations of the anomalous magnetic moment are improving rapidly
 and will be competitive with the R-ratio within the timescale of the Fermilab $g-2$ experiment.
As the LQCD data become more precise, methods such as that in Ref.~\cite{Blum:2018mom}
 can be used to combine the most precise parts of both R-ratio and lattice data to
 achieve an even smaller total uncertainty and to reduce the dependence on conflicting R-ratio data sets.
In addition, the lattice QCD calculations may be useful to help resolve these
 tensions.

In these proceedings, we discuss a recent calculation using LQCD techniques to improve
 the estimate of the HVP contribution to $(g-2)_\mu$.
This calculation is performed with 2+1 flavor sea and valence M\"obius domain wall fermions
 at physical $M_\pi$.
 In particular, we use an exclusive study with a hybrid of distillation \cite{Peardon:2009gh} and A2A \cite{Blum:2018mom} techniques
 to access the low-energy, long-distance
 tail of the local vector correlation function to provide additional constraint on the
 correlation function, reducing the large statistical uncertainty from that region.
We also compute $4\pi$ correlation functions to estimate the uncertainty
 associated with neglect of many-particle excited states on the HVP.
The distillation/A2A study is applied in conjunction with the improved bounding method \cite{Lehner:2018KEK},
 which uses the information
 garnered from the exclusive study to estimate the remaining contribution to the correlation
 function from neglected higher-energy excitations.
With these two improvements, we can achieve approximately a factor of 6 improvement
 in the uncertainty of the HVP estimate for one of our most precise ensembles.

\section{Analysis Method}
\label{sec:analysis}

The large correlator basis makes the
 Generalized Eigenvalue Problem~(GEVP)~\cite{Blossier:2009kd,Bulava:2014vua}
 an appealing strategy for computing the spectrum of states
 and overlap factors of each state in the correlation functions.
To this end, a matrix of correlation functions is constructed from considering a set of $N$
 operators and all cross-combinations of pairs of operators,
\begin{equation}
 \braket{ {\cal O}^{\dagger}_i(t) {\cal O}_j(0)} = C_{ij} (t)
 = \sum_n^\infty \bra{\Omega} {\cal O}^{\dagger}_i \ket{n}
 \bra{n} {\cal O}_j \ket{\Omega}
 e^{-E_n t}
 \,.
 \label{eq:correlator}
\end{equation}
We determine values for $\vert \bra{n} {\cal O}_j \ket{\Omega} \vert$ and $E_n$ from the GEVP, and can therefore partial reconstruct the correlation function
 up to state $M\leq N$.  We call this truncated correlation function $C^M$.

\begin{figure}[h]
\centering
 \begin{tabular}{cc}
 \includegraphics[width=0.47\textwidth]{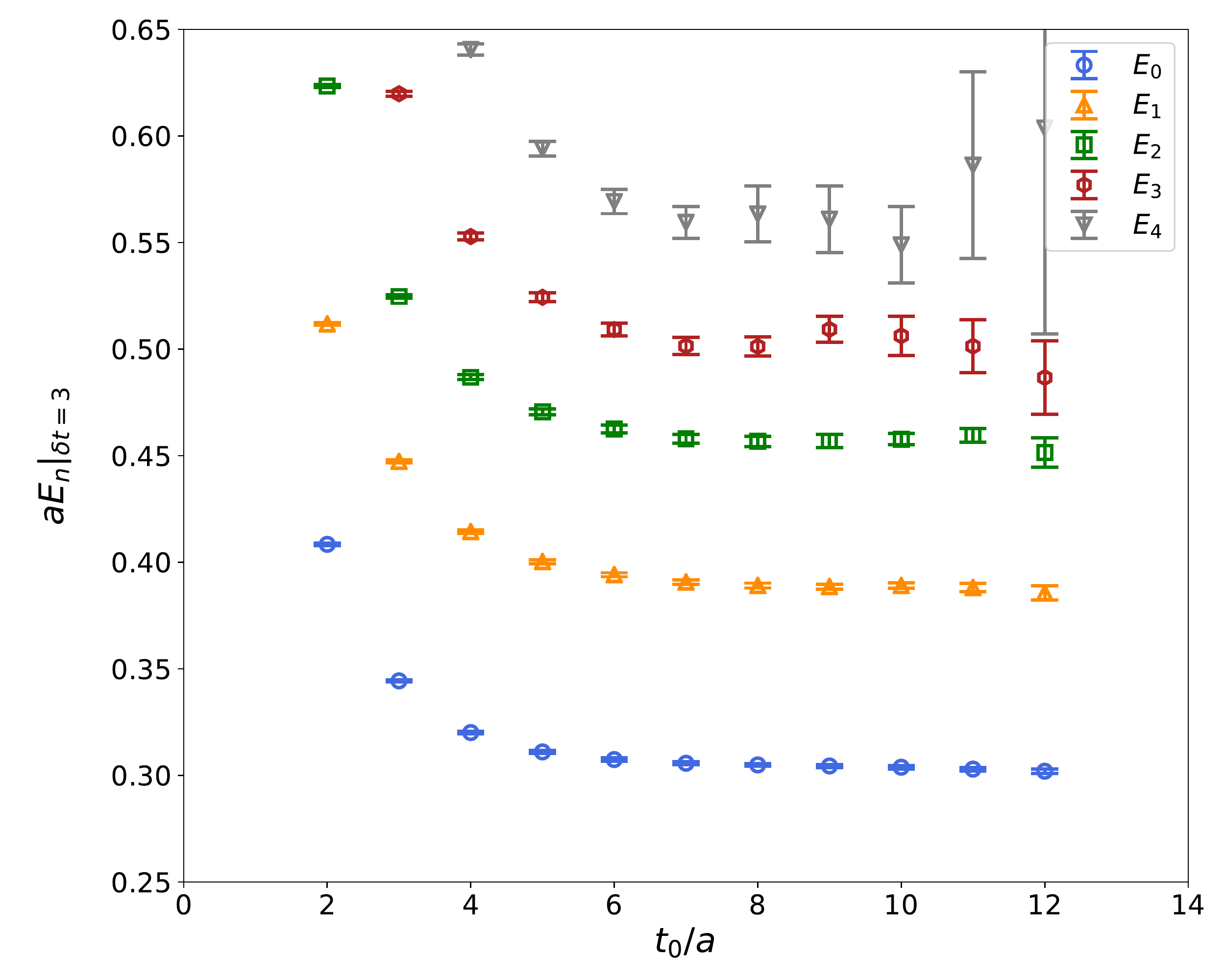} &
 \includegraphics[width=0.47\textwidth]{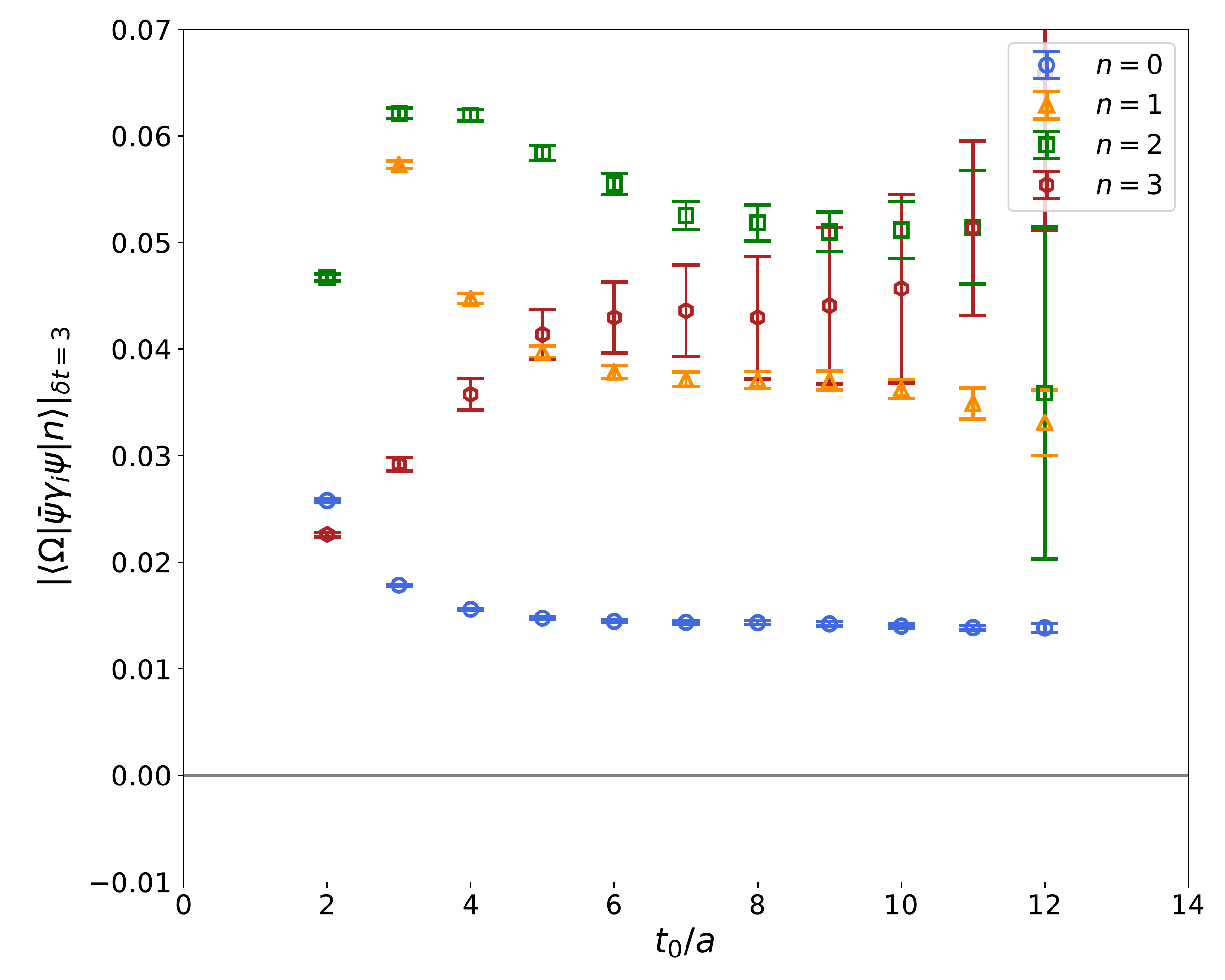} 
 \end{tabular}
 \caption{
 Left: spectrum obtained from solving the GEVP for a $C(t_0)C^{-1}(t_0+\delta t)$ for fixed $\delta t$.
 Small values of $t_0$ are subject to contamination from excited states,
  which is observed as an exponential approach to the asymptotic value at large $t_0$.
 Right: overlap of the states with the local vector current operator.
 Both left and right plots should asymptote to the true value in the large $t_0$ limit.
 In this figure, the GEVP is solved for a 6-operator basis and $\delta t=3$ is used.
 \label{fig:gevp2pi}
 }
\end{figure}

\begin{figure}[h]
\centering
 \begin{tabular}{cc}
 \includegraphics[width=0.47\textwidth]{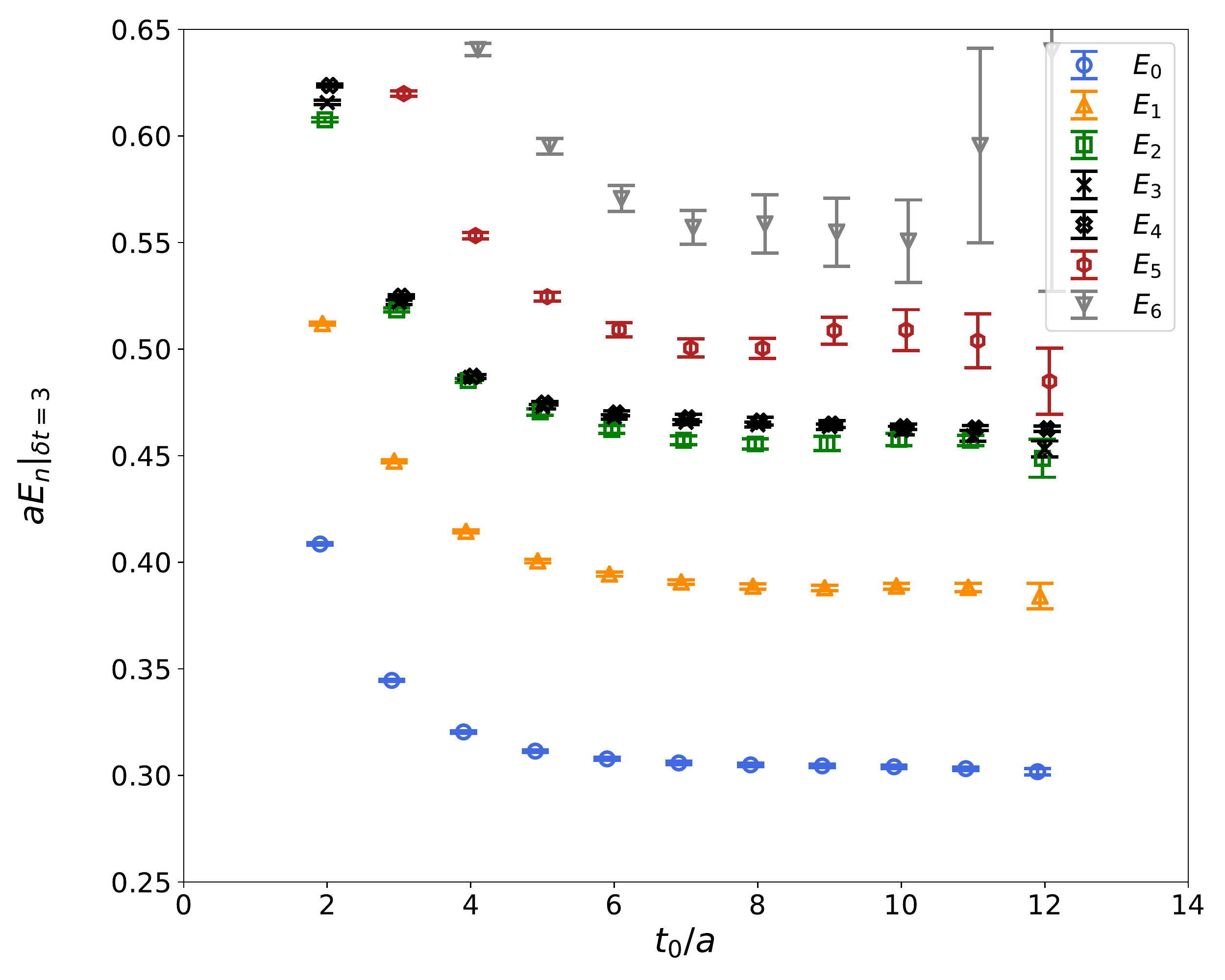} &
 \includegraphics[width=0.47\textwidth]{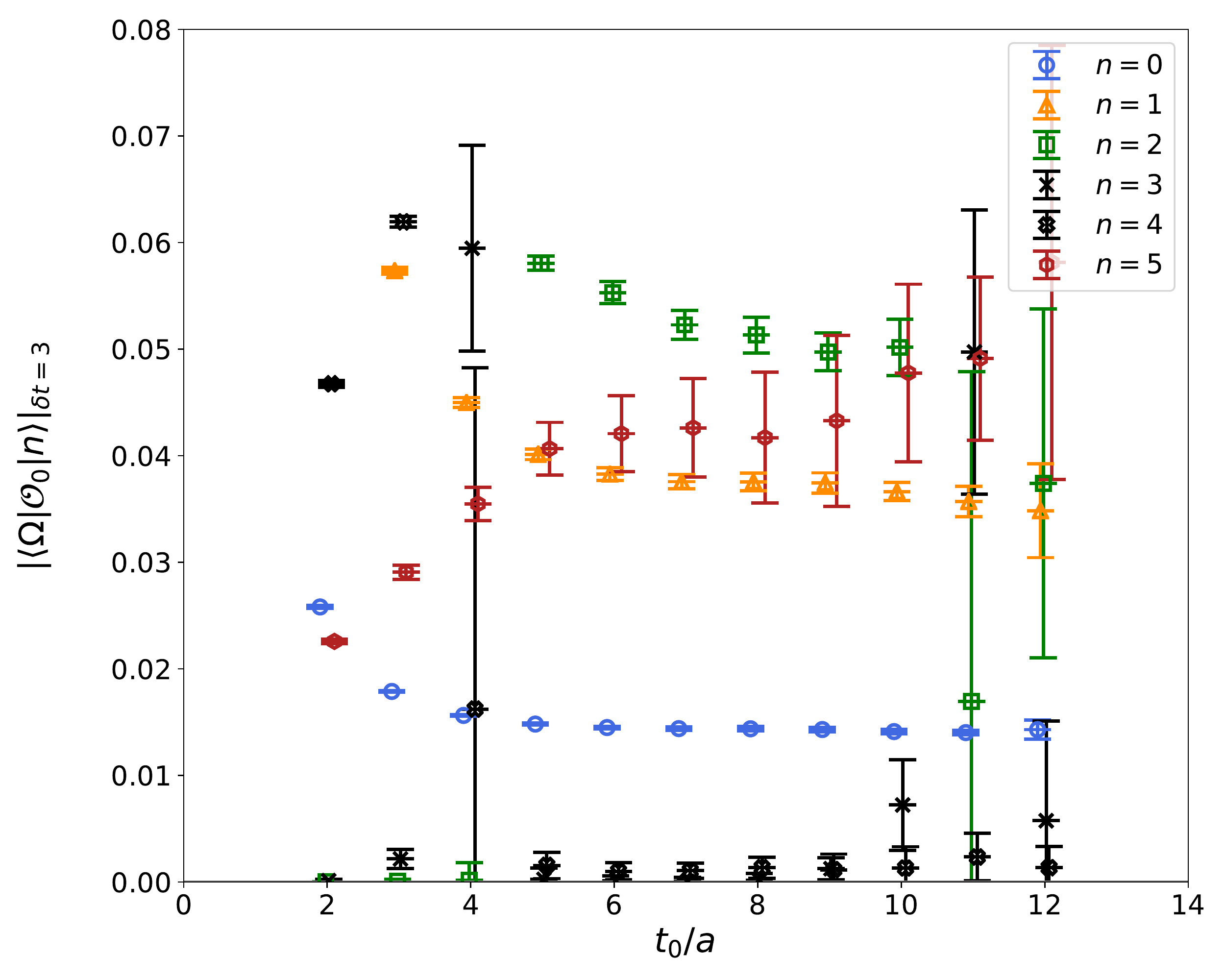} 
 \end{tabular}
 \caption{
 Same as Fig.~\ref{fig:gevp2pi}, except for a GEVP with 2 additional $4\pi$ operators.
 The new states that appear as a result of the addition of the $4\pi$ operators
  are denoted as black crosses, while the colors and symbols for the other states are 
  kept consistent with Fig.~\ref{fig:gevp2pi}.
 Even with the additional operators, the spectrum and overlaps
  are very strongly correlated with those without the additional operators.
 The large statistical error for $t_0=4$ in the right plot is associated with
  overlapping error bars, which results in eigenvalues that are not sorted properly.
 \label{fig:gevp4pi}
 }
\end{figure}

In this project, operators are constructed in the $I=1$ isospin representation,
 with $P$-wave spin channel to impact upon HVP$_{\mu}$.
First, we define a shorthand for a one-pion operator (ignoring isospin),
\begin{equation}
 {\cal O}_\pi^{p_\pi} = \sum_{xyz}
 \bar{\psi}(x) f(x-z) e^{-i \vec{p}_{\pi}\cdot\vec{z}} \gamma_5 f(z-y) \psi(y) \,,
\end{equation}
 where $f(x)$ is a kernel function representing the distillation smearing.
Three main choices of operators are used
 (ignoring isospin and momentum averaging for brevity):
\\[.5em]
\begin{tabular}{ll}
\quad $\bullet$ Local vector current   & $\sim \sum_x \bar{\psi}(x) \gamma_\mu \psi(x)$, $\mu\in\{1,2,3\}$\\[.5em]
\quad $\bullet$ Smeared vector current & $\sim \sum_{xyz} \bar{\psi}(x) f(x-z) \gamma_\mu f(z-y) \psi(y)$\\[.5em]
\quad $\bullet$ $2\pi$ operators
 & $\sim {\cal O}_\pi^{p_\pi} {\cal O}_\pi^{-p_\pi}$
\\
\end{tabular}\\[.5em]
with $\vec{p}_{\pi} \in \frac{2\pi}{L}\times \{ (1,0,0),(1,1,0),(1,1,1),(2,0,0) \}$
 for the $2\pi$ operators.
The results for applying the GEVP to this set of six operators is shown in
 Fig.~\ref{fig:gevp2pi}.
From these GEVP results, we can determine the spectrum for the first five
 states and overlaps for the first four states reliably.
The fifth spectrum energy is shown to demonstrate control over the next exponential
 in the sum.
We also test two $4\pi$ operators
 with $\vec{p}_{\pi} = \frac{2\pi}{L}\times (1,0,0)$:
\begin{itemize}
 \item $4\pi$ operators
 $\sim {\cal O}_\pi^{p_\pi} {\cal O}_\pi^{-p_\pi} {\cal O}_\pi^{0} {\cal O}_\pi^{0} $ \,,
\end{itemize}
in a world-first computation of $4\pi$ correlation functions in the $I=1$ channel.
The addition of these two operators is shown in Fig.~\ref{fig:gevp4pi}.
Adding the $4\pi$ operators does not appreciably change the GEVP results for the
 $2\pi$ states, as expected from phenomenology.
The remaining analysis will be carried out without the $4\pi$ operators in the basis.
The $3\pi$ states in this isospin channel are not relevant for this study,
 since they are required by parity symmetry to have large momenta that put them at
 energies above $1~{\rm GeV}$, and also because they exist only as lattice artifacts
 due to Wick symmetrization in the continuum and infinite volume limits.

\section{Results}
\label{sec:results}

The HVP contribution to the anomalous magnetic moment could be computed by applying
 the Bernecker-Meyer kernel in the time-momentum representation,
\begin{equation}
 a_\mu^{HVP} = \sum_{t=0}^{T/2} w_t C(t) \,,
 \label{eq:hvpcontrib}
\end{equation}
 with $C(t)$ from Eq.~\ref{eq:correlator} for the local vector current operator
 and $w_t$ from Ref.~\cite{Bernecker:2011gh}.

Rather than applying Eq.~\ref{eq:hvpcontrib} as is,
 the HVP contribution is computed by applying the bounding method,
 which employs a strict upper and lower bound on the correlation function~\cite{Lehner:2016BNL,BMW:2017}.
The bounding method is implemented as
\begin{equation}
 \widetilde{C}(t;t_{\text{max}},E) = \left\{
 \begin{array}{ll}
 C(t)             & t    < t_{\text{max}} \\[.5em]
 C(t_{\text{max}}) e^{-E(t-t_{\text{max}})} & t \geq t_{\text{max}}
 \end{array} \right.
 \label{eq:bounding}
\end{equation}
 where the exponential energy parameter $E$ is chosen separately for the upper and lower bound.
A systematic uncertainty is taken for the difference between upper and lower bounds,
 and optimal value of $t_{\text{max}}$ is chosen to minimize the uncertainty on $a_\mu^{HVP}$
 by balancing the systematic from low timeslices with
 the statistical uncertainty from large timeslices.

\begin{figure}[h]
\centering
 \begin{tabular}{cc}
 \includegraphics[width=0.47\textwidth]{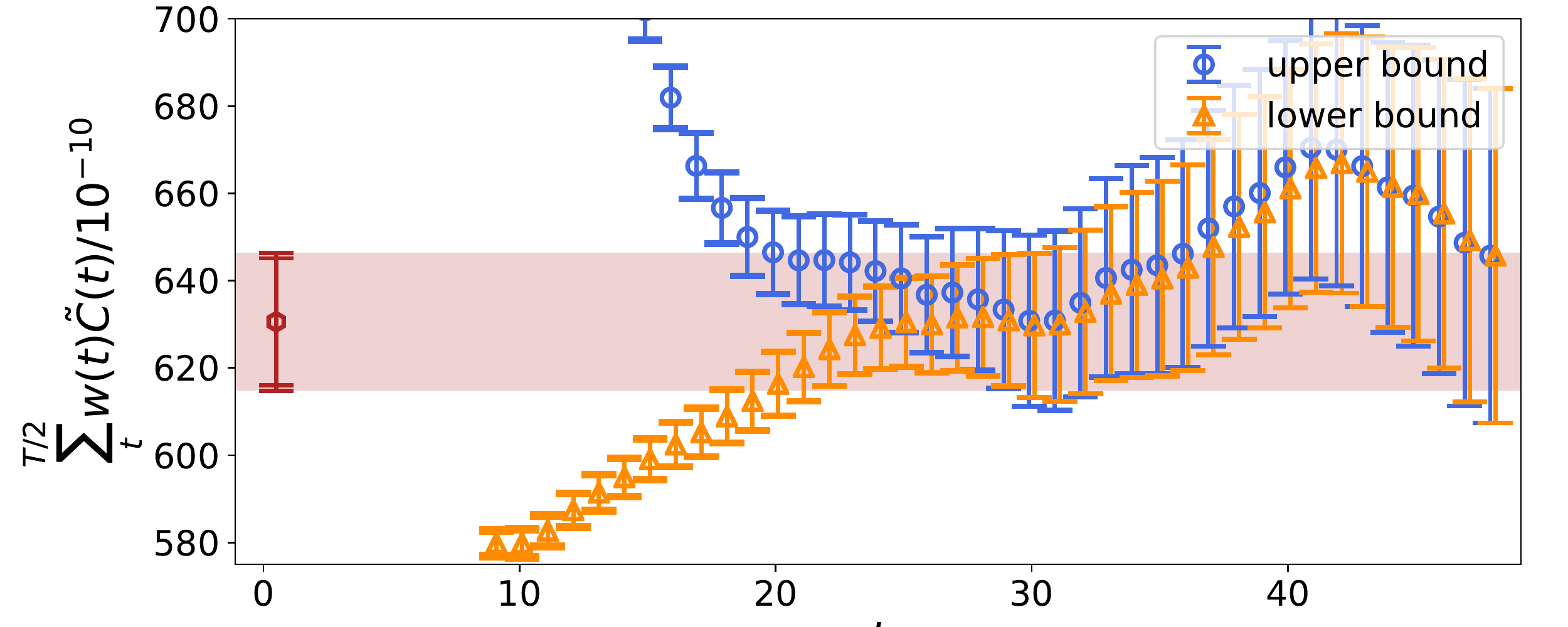} &
 \includegraphics[width=0.47\textwidth]{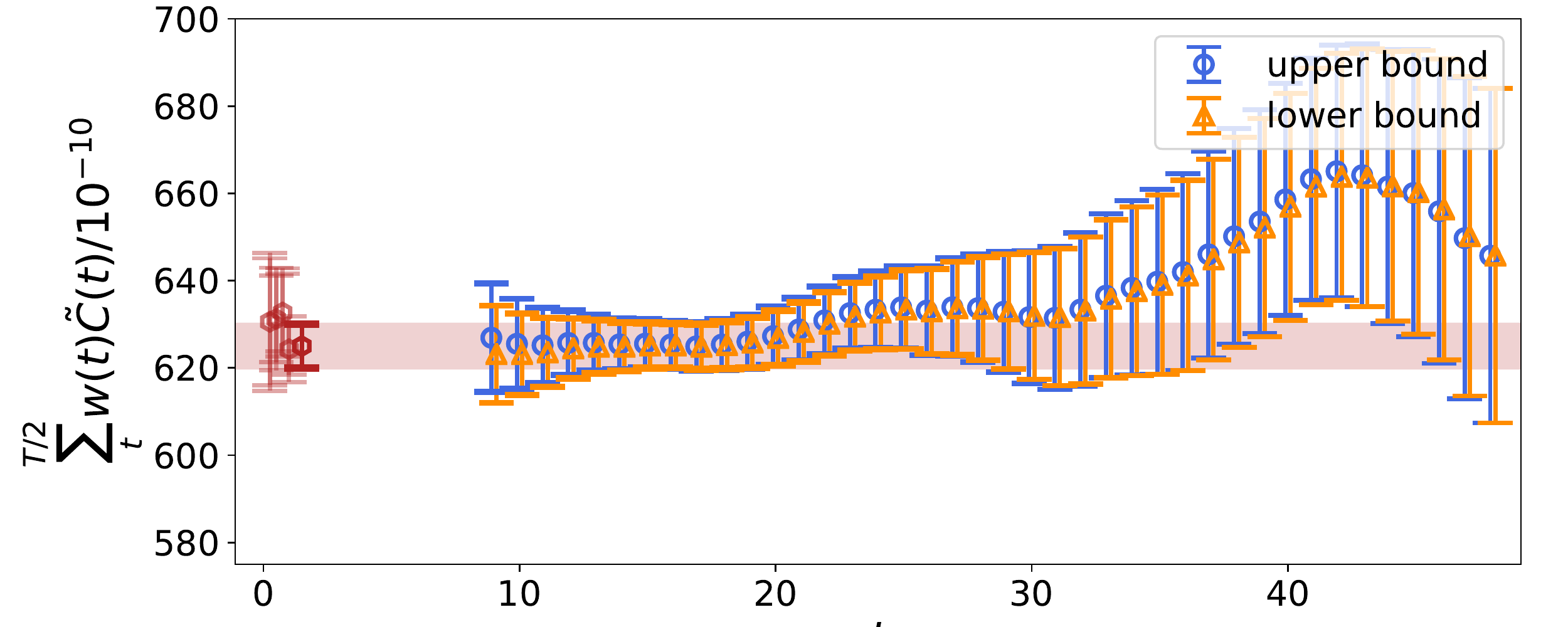} 
 \end{tabular}
 \caption{
 Bounding method applied to the local vector current correlation function.
 The upper bound and lower bound are applied as a function of $t_{\text{max}}/a$
  along the horizontal axis.
 The vertical axis is $a_\mu^{HVP}$ in units of $10^{-10}$ as measured from summing up
  the timeslices on the lattice.
 The left plot shows the bounding method with no improvement,
  and the right plot shows the improved bounding method with a 4-state reconstruction applied.
 As more states are added to the reconstruction, the upper and lower bounds converge
  at shorter $t_{\text{max}}$.
 The red point on the left side of both plots shows the optimal value obtained by picking
  the timeslice that minimizes the uncertainty (numerical values given in the text).
 In the right plot, the intermediate values from the $N$-state reconstructions with $N<4$
  are shown as well.
 \label{fig:bounding}
 }
\end{figure}

The bounding exponentials $E$ in Eq.~\ref{eq:bounding}
 are determined from known information about the correlation function.
The upper bound uses $E \leq E_0$, where $E_0$ is the lowest state in spectrum,
 which is determined from the GEVP spectrum.
The lower bound is determined from the local effective mass,
 $E \geq {\rm log}[\frac{C(t_{\text{max}})}{C(t_{\text{max}}+1)}]$.
These are strict bounds, but any choice of $E$ greater (lower) than these values
 for the lower (upper) bound is also valid.
The results for applying the procedure of Eq.~\ref{eq:bounding} for different values of
 $t_{\text{max}}$ are shown in the left panel of Fig.~\ref{fig:bounding}.

\begin{figure}[h]
\centering
 \begin{tabular}{cc}
 \includegraphics[width=0.47\textwidth]{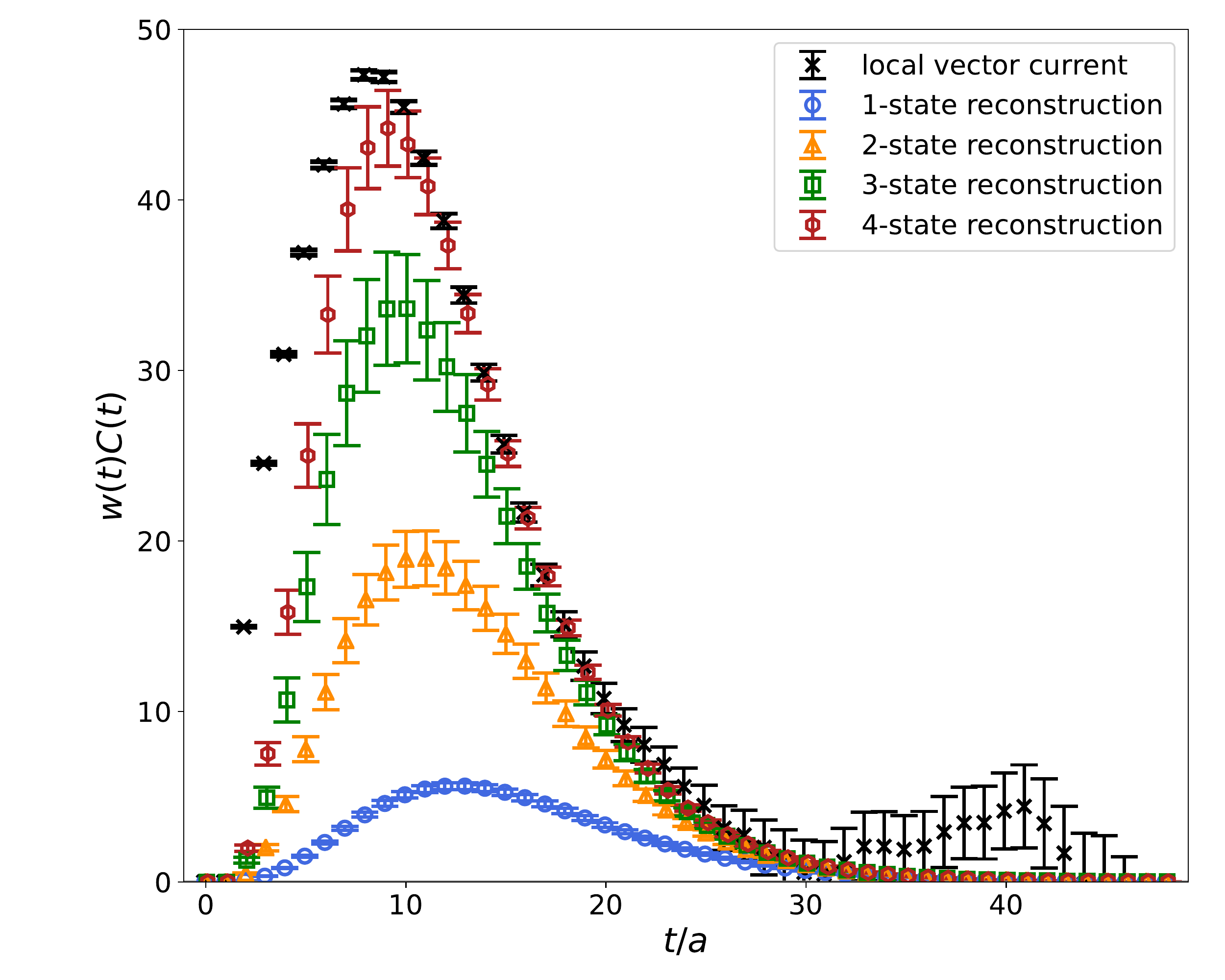} &
 \includegraphics[width=0.47\textwidth]{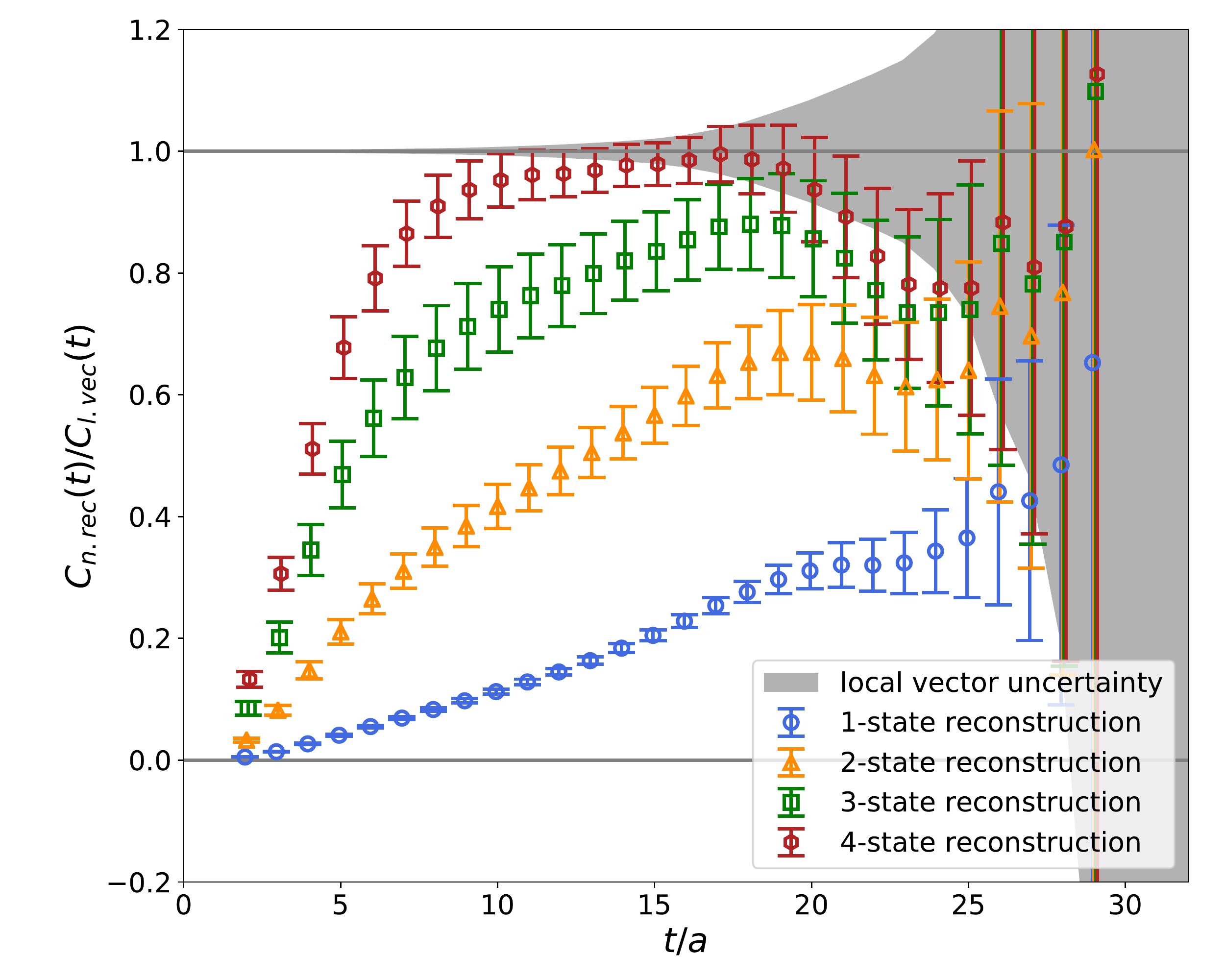} 
 \end{tabular}
 \caption{
 Left: Integrand of $a_\mu^{HVP}$ plotted as a function of $t/a$.
 The local vector current correlation function by itself is plotted as black crosses,
  and the $N$-state reconstruction obtained from the GEVP are shown in colors.
 As more states are added to the correlation function reconstruction, the resulting
  curve shape matches the local vector current down to shorter distance.
 Right: Ratio of the $N$-state reconstructions normalized by the local vector current
  correlation function.
 The uncertainty on the local vector current correlation function is denoted by the
  gray band.
 As more states are added, the ratio of reconstruction over local vector current approaches 1,
  and the 4-state reconstruction gives a reconstruction consistent with the local vector
  current to within $1\sigma$ after about $t/a=10$.
 \label{fig:reconst}
 }
\end{figure}

Applying the information from the full GEVP spectrum study can be used to
 improve the bounding method~\cite{Lehner:2018KEK}.
The precision of the final result will depend on how well the parameters of the
 correlation function $C(t)$ can be reconstructed from the GEVP in the previous section.
The left plot of Fig.~\ref{fig:reconst} shows the overall contributions from the lowest 4 states,
 and the right plot shows the ratio of the reconstruction over the local vector current.
The 4-state reconstruction gives a good description of the correlation function down to
 around timeslice 10, or about $1~{\rm fm}$.

To use the reconstruction from the GEVP results,
 rather than applying the bounding to the raw correlation function, $C(t)$ in Eq.~\ref{eq:bounding}
 is replaced by a new correlation function where the reconstruction of the lowest $M$ states
 is subtracted away from the full local vector correlation function,
\begin{equation}
 C(t)\to C(t)-C^M(t) \,,
 \label{eq:boundingimprovement}
\end{equation}
for $t> t_{\text{max}}$ with $C^M(t)$ defined below Eq.~\ref{eq:correlator}.
The values for the lowest $M$ states are taken from the GEVP.
This subtracted correlation function will have a faster falloff for the upper bound,
 which now takes $E \to E_{M+1}$ as the lowest state in the spectrum,
 and so will converge with the lower bound at smaller $t_{\text{max}}$.
Here, $E_{M+1}$ is the next state in the spectrum and can also be estimated from the GEVP.
The anomalous magnetic moment is obtained from the relation
\begin{equation}
 a_\mu^{HVP} =
 \sum_{t=0}^{t_{\text{max}}} w_t C(t)
 + \sum_{t=t_{\text{max}}+1}^{T/2} w_t \widetilde{C}(t;t_{\text{max}},E)
 + \sum_{t=t_{\text{max}}+1}^{T/2} w_t C^M(t)\,,
\end{equation}
 where the first term on the RHS is the normal HVP kernel (with no subtraction)
 summed up to $t_{\text{max}}$,
 the second term is bounded by the procedure in Eq.~\ref{eq:bounding}
 after applying the subtraction in Eq.~\ref{eq:boundingimprovement},
 and the third term is the reconstructed correlator fixed by the GEVP fit.
 The results of the improved bounding procedure with a 4-state reconstruction implemented
 in Eq.~\ref{eq:boundingimprovement} are shown in the right panel of Fig.~\ref{fig:bounding}.

With the techniques applied here, a significant improvement in the statistical precision
 can be achieved.
Summing the local vector correlation function directly with Eq.~\ref{eq:hvpcontrib}
 yields the value $a_\mu^{HVP} = 646(38)\times10^{-10}$.
Applying the bounding method without improvement decreases the uncertainty by a factor of 2,
 giving $a_\mu^{HVP} = 631(16)\times10^{-10}$,
 and improvement of the bounding method by applying a four state reconstruction
 results in an additional factor of 3 improvement to give a result of $625.0(5.4)\times10^{-10}$.
 These two values correspond to the values obtained from optimizing the bounding method
 in Fig.~\ref{fig:bounding} for the left and right plots, respectively.

\section{Outlook and Conclusions}
\label{sec:conclusions}

In these proceedings, we have demonstrated techniques for improvement of a lattice QCD-only
 calculation of the HVP contribution to the muon $g-2$.
The exclusive study using distillation/A2A allows for control of the long-distance tail of
 the local vector current correlation function.
The large statistical uncertainty of this tail is replaced by the significantly smaller
 systematic uncertainty associated with fitting systematic constraints.
The bounding method provides a robust way to estimate the systematic effects of the
 reconstruction on the large-time correlation function.
The reconstruction of the low-energy spectrum and overlaps of the local vector current
 correlation function is also used to improve the bounding method,
 garnering an additional factor gain in the precision.
With these techniques applied, the precision on the HVP contribution to the muon
 anomalous magnetic moment is improved by about a factor of 6,
 from $a_\mu^{HVP} = 646(38)\times10^{-10}$ to $a_\mu^{HVP} = 625.0(5.4)\times10^{-10}$
 on one ensemble.
We have also computed the contribution from the lowest $4\pi$ states
 in the vector current correlator and found these contributions to be negligible.

The techniques used here were formerly applied in Ref.~\cite{Lehner:2018KEK}
 to the HVP on two different lattice volumes
 and found to be precise enough to explicitly resolve the finite volume contributions
 at physical $M_\pi$.
We are currently working on computing the HVP contribution on another ensemble
 closer to the continuum limit.
This ensemble, combined with the strategies demonstrated in these proceedings,
 can be used to greatly improve the precision on the HVP contribution from
 $14\times10^{-10}$ down to $5\times10^{-10}$, with an additional improvement
 after the full set of systematic improvements are included.
With these improvements in estimates of the uncertainty,
 it is foreseeable that the precision on the HVP from theory will be able to match
 the experiment by the time the Fermilab $g-2$ experiment reaches its final precision.

\section{Acknowledgements}

We acknowledge Jozef Dudek, Christopher Thomas, David Wilson, Antoni Woss, and our colleagues in the RBC \& UKQCD collaborations
for interesting discussions.
This work used computer resources from the
 Extreme Science and Engineering Discovery Environment (XSEDE),
 which is supported by National Science Foundation grant number ACI-1548562.
This research used computational resources of the HPCI system provided by
 the University of Tokyo through the HPCI System Research Project (Project ID:hp180151, hp190137).
We gratefully acknowledge computer resources at the Oakforest-PACS supercomputer system at Tokyo University and at
the Theta supercomputer at Argonne National Laboratory.
We are grateful for computing resources provided through USQCD and the BNL SDCC at clusters at
Brookhaven National Lab and Jefferson Lab.

\def\bibspace{-0.18cm}
\vspace{\bibspace}

\end{document}